\documentclass[twocolumn,10pt]{revtex4}%
\usepackage[paperwidth=210mm,paperheight=297mm,centering,hmargin=2cm,vmargin=2.5cm]{geometry}
\usepackage{amsfonts}
\usepackage{amsmath}
\usepackage{amssymb}
\usepackage{bm}
\usepackage{graphicx}
\usepackage{color}
\usepackage{soul}
\usepackage{epsfig}%
\usepackage[dvipsnames]{xcolor} 
  \definecolor{bleu_cite}{RGB}{0,0,255}
\usepackage[colorlinks=true,allcolors = black,citecolor=bleu_cite,  ]{hyperref} 
\usepackage{natbib}

\usepackage{siunitx}

\begin{document}
\title{Key role of the moir\'{e} potential for the quasi-condensation of interlayer excitons in van der Waals heterostructures}

\author{Camille Lagoin  and Fran\c{c}ois Dubin} 
\affiliation{Institut des Nanosciences de Paris, CNRS and Sorbonne Universit{\'e}, 4 pl. Jussieu,
75005 Paris, France}

\begin{abstract}
 Interlayer excitons confined in bilayer heterostructures of transition metal dichalcogenides (TMDs) offer a promising route to implement two-dimensional dipolar superfluids. Here, we study the experimental conditions necessary for the realisation of such collective state. Particularly, we show that the moir\'{e} potential inherent to TMD bilayers yields an exponential increase of the excitons effective mass. To allow for exciton superfluidity at sizeable temperatures it is then necessary to intercalate a high-$\kappa$ dielectric between the monolayers confining electrons and holes. Thus the moir\'{e} lattice depth is sufficiently weak for a superfluid phase to theoretically emerge below a critical temperature of around 10 K. Importantly, for realistic experimental parameters  interlayer excitons quasi-condense in a state with finite momentum, so that the superfluid is optically inactive and flows spontaneously. 

\end{abstract}

\maketitle

\textit{Introduction} $-$ Since seminal theoretical predictions were formulated in the 1960's \cite{Blatt}, semiconductor excitons have been at the focus of a long-lasting research aiming at Bose-Einstein condensation and its related superfluidity. Triggered by the proposal of Lozovik and Yudson \cite{Lozovik_Yudson}, two-dimensional excitons confined in semiconductor bilayers have turned as the most favorable candidates to reach this objective. Such bilayer excitons are characterised by the separation enforced between their electron and hole constituents, which are each confined in a different layer. As a result they exhibit a long lifetime, \textit{a priori} sufficient for cold gases to reach thermodynamically quantum degeneracy \cite{Ivanov_2004}. 

Relying on unmatched material quality thanks to molecular beam epitaxy, studies of bilayer excitons confined in GaAs coupled quantum wells have traced the path for engineering quasi-condensates \cite{review}. Indeed, in GaAs bilayer excitons are studied in a homogeneously broadened regime \cite{Dang_2020} where the fingerprints of quasi-condensates are found in the photoluminescence radiated by optically bright states. Thus one observes algebraically decaying temporal coherence \cite{Dang_2020} combined to quasi long-range spatial coherence \cite{Dang_2019,Anankine_2017}. Importantly, these signatures are consistent with a Berezinskii-Kosterlitz-Thouless transition \cite{Berenzinskii,KT} expected for this two-dimensional geometry. 

In GaAs, bilayer excitons have a binding energy of a few meV that limits the maximum exciton density to less than 10$^{11}$ cm$^{-2}$\textcolor{black}{\cite{Ivanov_2004,review,Dang_2020,Dang_2019,Anankine_2017}}. Therefore, the critical temperature for their quasi-condensation is bound to about 1K. This limitation does not arm fundamental studies but in practice precludes device applications for quantum technologies. In fact, these require much higher operating temperatures, which may be accessed by interfacing atomic layers of transition metal dichalcogenides (TMDs)  \cite{Fogler_2014}, in so-called van der Waals heterostructures \cite{Geim_2013,Xu_2018,Tran_2020}. Electrons and holes thus have minimum energy states lying in a different layer, thereby implementing bilayer (interlayer) excitons. These exhibit binding energies up to 50 times greater than in GaAs \cite{Nature_Photonics_2016}. As a result, they are possibly stable at high densities, above 10$^{12}$ cm$^{-2}$\cite{Fogler_2014}, and collective quantum phenomena potentially emerge below higher critical temperatures. \textcolor{black}{Let us also note that in TMDs a spatial separation between opposite charge carriers is either ensured by the difference between the energy gap and work functions of two mono-layers for hetero-bilayers \cite{Xu_2018}, e.g. MoSe$_2$/WSe$_2$, whereas in homo-bilayers an external electrical polarisation is necessary as for GaAs bilayers \cite{review,Dang_2019,Anankine_2017}}

\textcolor{black}{Importantly, in TMDs bilayer excitons are inherently subject to a spatially modulated potential:} the so-called moir\'{e} potential that results from a non-zero twist angle between interfaced mono-layers, and/or a mismatch between their lattice constants \cite{Xu_2018}. Thus, the electronic bandgap varies spatially, with a period typically around 10-30 nm and an amplitude governed by the exact heterostructure design. Precisely, when two distinct monolayers are interfaced directly the moir\'{e} potential has been measured as large as 150 meV \cite{Zhang_2017}. On the other hand, when one or two \textcolor{black}{ hexagonal boron nitride (hBN)} monolayers are intercalated between two TMDs, the moir\'{e} potential has an amplitude that is expected around $5-10$ meV \cite{Imamoglu_2020}. In general, the moir\'{e} potential offers formidable opportunities to spatially arrange electronic carriers, at the nanoscale: evidence for electron crystallisation in the moir\'{e} lattice were reported recently \cite{Imamoglu_2020,Wign_1,Wign_2,Wign_3}, as well as the localisation of interlayer excitons \cite{moire_1,moire_2,moire_3,moire_4}.

In this work, we highlight that the moir\'{e} potential governs the parameter space where the quasi-condensation of interlayer excitons is accessible. By directly solving the Schroedinger equation, we first emphasize that interlayer excitons experience an exponential increase of their effective mass due to the moir\'{e} potential. In the regime where the latter has a large depth, the excitons effective mass is increased by orders of magnitude, \textcolor{black}{so that the critical temperature for quantum degeneracy is too small to be accessed by cryogenic techniques}. On the other hand, for weak moir\'{e} potential depths we determine the experimental parameters, e.g. the period and the depth of the moir\'{e} lattice potential, where exciton superfluidity may be accessed. For that, we rely on a mean-field treatment of the Bose-Hubbard model and show that exciton superfluidity is favourable when the moir\'{e} potential depth is less than 10 meV, below a critical temperature of around 10 K. We then discuss our findings with regards to the experimental state-of-the-art, as well as accessible experimental signatures for exciton superfluids.

\textit{Moir\'{e} potential and interlayer excitons luminescence} $-$ For simplicity, let us consider an heterobilayer such as WX$_2$/MoY$_2$, each layer confining electrons or holes constituting interlayer excitons. Stacking two of such monolayers yields a moir\'{e} lattice with a period given by $a_m\sim a_0/\sqrt{\theta^2+\delta^2}$ where  $\theta$ denotes the twist angle between the layers and $\delta$=$|a_0-a_1|$/$a_0$, $a_{0,1}$ being the layer lattice constants. Interestingly, the lattice mismatch \textcolor{black}{$\delta$}  is greatly reduced when the two chalcogen atoms X and Y are identical ($\delta\sim0.1\%$), whereas it can reach a few $\%$ otherwise \textcolor{black}{\cite{apl_1991}}. Figure 1.a illustrates the moir\'{e} lattice by depicting the real space arrangement of atoms for two monolayers stacked with a small twist angle, neglecting $\delta$.

\begin{figure}[!ht]
  \includegraphics[width=\linewidth]{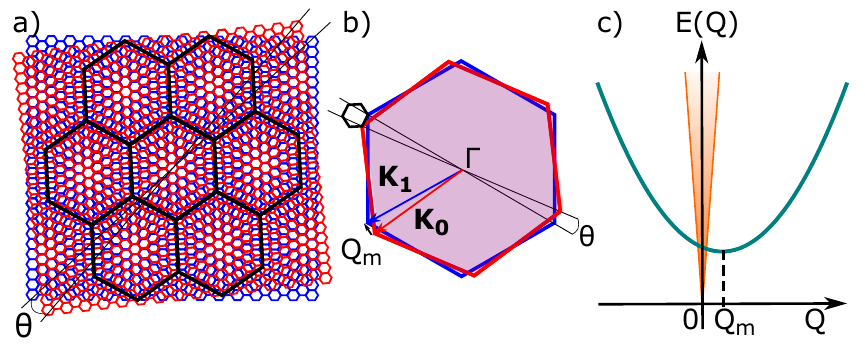}
  \caption{(a) Real space arrangement of atoms for two monolayers, red and blue respectively, stacked with a small twist angle $\theta$, neglecting the lattice mismatch. The resulting moir\'{e} lattice is drawn in black. (b) Brillouin zones of the two monolayers together with the moir\'{e} Brillouin zone (black) resulting from the non-vanishing twist angle $\theta$. The latter controls the \textcolor{black}{wave-vector} separation $Q_m$ between the corners of the moir\'{e} Brillouin zone. (c) Dispersion of interlayer excitons for a non-vanishing twist angle (green) together with the photon dispersion (orange). Their intersection marks the optically active region of the exciton band.}
  \label{fig:fig1}
\end{figure}

An important consequence of the moir\'{e} lattice regards the optical activity of interlayer excitons. Indeed, these are made of electrons and holes each confined in a different layer, and by discarding $\delta$ for simplicity\textcolor{black}{,} we note that a finite twist angle between the monolayers implies that their reciprocal lattices experience a net relative rotation (Fig.1.b). As a result, the extrema of the valence and conduction bands of each monolayer are not aligned in reciprocal space. Thus, interlayer excitons have a lowest energy state lying at $Q_m$ from the optically active region \textcolor{black}{with quasi-vanishing in-plane wave-vector} (Fig.1.c). For a WSe$_2$/MoSe$_2$ hetero-bilayer with a typical twist angle $\theta\sim$1$^\circ$ we deduce that $Q_m\sim$ 130 $\mu$m$^{-1}$. Then, a quasi-condensate of interlayer excitons, necessarily occurring due to a macroscopic occupation of the lowest energy state \cite{Monique_2015}, does not radiate any photoluminescence. Instead, it carries a finite momentum $\hbar Q_m$, \textcolor{black}{$\hbar$ being the reduced Planck constant}. Let us finally note  that engineering an optically bright quasi-condensate, i.e. with around zero momentum, requires a maximum twist $\theta\sim$ 0.2$^\circ$  considering a WSe$_2$/MoSe$_2$ hetero-bilayer.

\textit{Exciton effective mass in a moir\'{e} potential}$-$ Besides inducing a relative shift of conduction and valence bands in reciprocal space, the moir\'{e} potential also leads to a modulation of the excitons potential energy in real space. Considering this modulation is necessary to accurately model the optical selection rules of interlayer excitons \cite{McDonald_2018,Yu_2017}. In the following we show that it is also necessary to carefully consider the depth of the moir\'{e} potential to accurately evaluate the excitons effective mass.

The hamiltonian for interlayer excitons exploring a moir\'{e} potential can be simply expressed as 
\begin{equation}
H = \frac{-\hbar^2}{2m}\frac{\partial^2}{\partial z^2} + s E_R sin^2(\textcolor{black}{q_m} z)
\end{equation}
by restricting the excitonic motion to one direction $z$ of the moir\'{e} lattice, \textcolor{black}{which associated wave vector reads in one dimension $q_m=\pi/a_m$}. In Eq.(1) $m$ denotes the exciton effective mass. \textcolor{black}{In the second term $sE_R=V_0$ provides the depth of the moir\'{e} lattice, with $E_R=\hbar^2q_m^2/2m$}. Of course, the hamiltonian (1) does not take into account the moir\'{e} potential in its full microscopic complexity. Nevertheless, Eq.(1) allows one to accurately extract the renormalisation of the effective mass \textcolor{black}{induced} by a lattice potential, as shown for ultra-cold atoms confined in square optical lattices  \cite{Stringari_2002}. Therefore we follow this approach here in the context of interlayer excitons exploring a triangular moir\'{e} potential.

Using the above hamiltonian we look for the solutions of the eigen equation $H \psi_p$=$E(p)\psi_p$,  \textcolor{black}{$p$ being the exciton quasi-momentum}. This turns into solving a Mathieu equation \cite{Mathieu} so that the solutions $\psi_p$ are Mathieu functions, having a period $a_m$ by definition. The excitons effective mass $m^*$, dressed by the moir\'{e} potential, is then deduced from the energy dispersion by setting $E(p)\sim E_0+p^2/2m^*$. Let us note that this matching is accurate for both weak and deep moir\'{e} lattices, although in the former case one needs to include on-site interactions into the hamiltonian in order to exactly deduce $m^*$ \cite{Stringari_2002}.

\begin{figure}[!ht]
  \includegraphics[width=\linewidth]{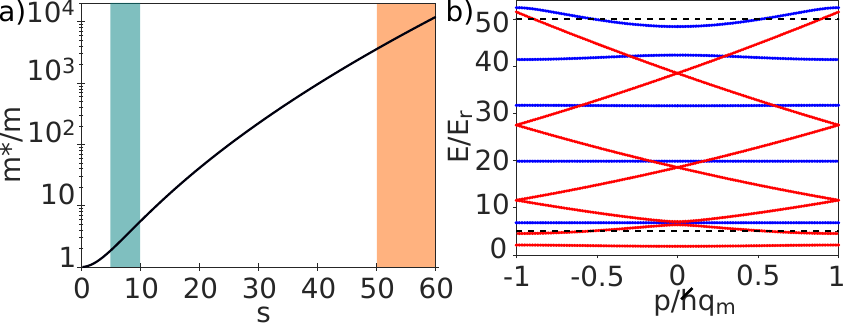}
  \caption{(a) Ratio between the excitons effective mass renormalised by the moir\'{e} lattice and the excitons bare effective mass, $m^*/m$, \textcolor{black}{ as a function of $s$ quantifying the depth of the moiré potential $V_0 $ in units of $E_R$}. The green and orange shaded areas mark the parameter range for bilayers with and without hBN spacer respectively. (b) Exciton energy bands in the moir\'{e} Brillouin zone for $s=5$ (red) and $s=50$ (blue). The black dashed lines display the corresponding values of the potential depth $s$. Energies are expressed in units of $E_R$ and the quasi-momentum $p$ in units of \textcolor{black}{$\hbar q_m$.}}
  \label{fig:fig2}
\end{figure}

Figure 2.a presents the ratio $m^*/m$ as a function of $s$. As expected, for vanishing moir\'{e} potentials ($s\sim0$) we find that $m^*\sim m$ whereas $m^*/m\sim$2500 for deep moir\'{e} lattices ($s\sim$50). In fact, Fig.2.a shows that $m^*$ exponentially increases with $s$. This universal scaling as a function of $s$ allows us to extract effective masses for heterostructures that are currently studied to explore the quasi-condensation of interlayer excitons. First, we consider a MoSe$_2$/WSe$_2$ heterobilayer as probed in Ref. \cite{Holleitner_2020}. In this device electrons and holes have effective masses $m_e\sim$ 0.49 $m_0$ and $m_h\sim$ 0.35 $m_0$ respectively, leading to $m=m_e+m_h\sim$ 0.84 $m_0$, $m_0$ denoting the free electron mass. For a typical 1$^{\circ}$ twist angle between stacked monolayers, the moir\'{e} lattice has a period $a_m\sim$ 19 nm leading to $E_R\sim$ 1.2 meV. The depth $V_0$ of the moir\'{e} lattice has not been measured for MoSe$_2$/WSe$_2$ heterobilayers, but for MoS$_2$/WSe$_2$ scanning tunneling microscopy has revealed that it is of around 150 meV \cite{Zang_2017}. Furthermore, DFT calculations have confirmed that $V_0\sim$ 110 meV for MoSe$_2$/WSe$_2$ \cite{McDonald_2018}. This implies that $s\sim$ 100 leading to $m^*/m\sim$ 10$^6$. Moreover, independent experiments \cite{Shan_2019} have reported studies of  a MoSe$_2$/WSe$_2$ bilayer when two monolayers of hBN separate the TMDs. For such device the depth of the moir\'{e} potential is highly screened by hBN so that $V_0$ can not exceed 5-10 meV \cite{Imamoglu_2020}, leading to $s\sim$ $5-10$ and $m^*/m\sim$ $1.5-5$. 

Having included the effect of the moir\'{e} lattice onto the effective mass of interlayer excitons we now deduce the critical temperatures where quantum degeneracy and superfluidity theoretically occur, $T_d$ and T$_{BKT}$ respectively. These are ruled by both the exciton density $n$ and their effective mass $m^*$. They read $
T_d = \frac{2\pi\hbar^2}{k_B m^*}\frac{n}{g}$ and $T_{BKT}\sim \frac{\pi}{2} \frac{\hbar^2}{k_B m^*}n_s$  \cite{Filinov_2010}, $k_B$ being the Boltzmann constant while $n_s\sim \alpha n/g$ denotes the superfluid density with $n$ the total density and $g$ the degeneracy of the lowest energy excitonic band. Taking into account the spin-orbit splitting of the conduction band, we set $g$=2 whereas for 10$^{10}\lesssim n\lesssim 10^{12}$ cm$^{-2}$ $\alpha$ ranges between 0.6 and 0.9 \cite{Filinov_2010}. Figure 3.a displays the variations of T$_{BKT}$ as a function of $s$ and $n$. For a MoSe$_2$/WSe$_2$ heterobilayer we strikingly note that T$_{BKT}$ is bound to the milli-Kelvin range, so is $T_d$. These magnitudes directly reflect the exponentially increased $m^*$. On the other hand, when a few hBN monolayers are intercalated between the two TMDs, Fig.3 shows that both $T_d$ and T$_{BKT}$ reach sizeable values, up to $T_d\sim$ 30 K and T$_{BKT}\sim$ 7 K when $n\sim$ 10$^{12}$ cm$^{-2}$ as shown in Fig 3.b.
\begin{figure}[!ht]
  \includegraphics[width=\linewidth]{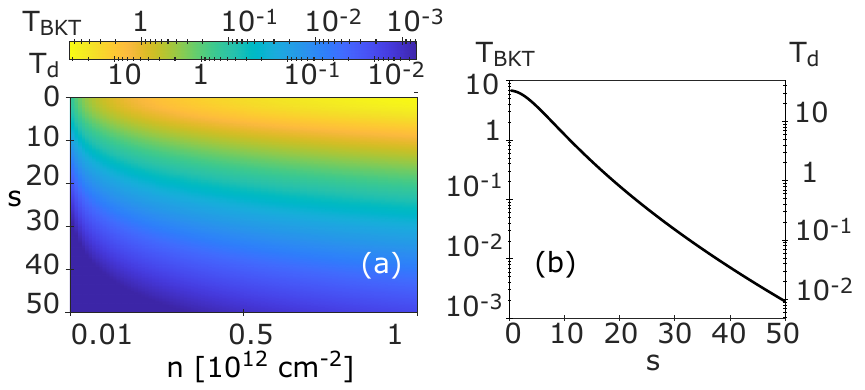}
  \caption{(a) Critical temperatures for quantum degeneracy $T_d$ and for the Berezinskii-Kosterlitz-Thouless crossover $T_{BKT}$, as a function of the exciton density $n$ and the depth of the moir\'{e} potential $s$ \textcolor{black}{which gives $V_0 $ in units of $E_R$}.	(b) $T_d$ and T$_{BKT}$ as a function of $s$ for an exciton density set to 10$^{12}$ cm$^{-2}$. Temperatures are displayed in logarithmic scale and Kelvin.}
  \label{fig:fig3}
\end{figure}

Since T$_{BKT}$ scales linearly with $n$ it is tempting to anticipate that high-temperature superfluids are accessible at high densities. However, we would like to point out that for two-dimension dipolar systems this assumption is not straightforward. Indeed, path integral Monte-Carlo calculations have shown that quasi-condensation of dipolar excitons is bound to a maximum density $n$, defined by $D\lesssim15$ where \textcolor{black}{$D=d^2/4\pi\epsilon a^3 E_0$, $d$ being the exciton electric dipole moment, $\epsilon=\epsilon_r\epsilon_0$ is the dielectric permittivity with $\epsilon_r=3$ if we consider that
the heterostructure is exposed to hBN and vacuum}, while $a=1/\sqrt{n}$ and $E_0=\hbar^2/m^* a^2$ \cite{Filinov_2010}. Thus, we deduce that quasi-condensation becomes inaccessible beyond a maximum density of around 2.5 10$^{12}$ cm$^{-2}$ leading to T$_{BKT}\sim$ 12 K, since in this situation the quasi-condensed fraction is reduced to around 60$\%$. Note that $n_s$ is maximised to about 90$\%$ for $D\sim$ 1, obtained for $n\sim$ 2 10$^{11}$ cm$^{-2}$ yielding T$_{BKT}\sim$ 2 K. Let us then stress that experiments with bilayer excitons in GaAs have actually confirmed the breakdown of a quasi-condensate for $D\gtrsim12$ \cite{Dang_2019}.

\textit{Mott insulator vs. superfluid of interlayer excitons}$-$ \textcolor{black}{In the following, we focus onto devices where hBN separates the monolayers confining electrons and holes, so that T$_{BKT}$ is maximised. However in the moir\'{e} lattice the emergence of exciton superfluidity is not necessarily favourable energetically below this estimated critical temperature.} This actually depends on the competition between the exciton interaction strength in the moiré lattice sites and the strength of exciton tunnelling between neighbouring sites. This competition is described by the Bose-Hubbard model \cite{Salomon_2010} predicting that at least two antagonist states compete in the quantum regime. These are namely the Mott insulating phase, characterised by a fixed number of particles in each lattice site, and a superfluid phase marked by quasi long-range order. The Bose-Hubbard hamiltonian usually reads 
\begin{equation}
\hat{H}_{BH} = -t \sum_{i,j} \hat{b}^{\dag}_i \hat{b}_j +\frac{U}{2}\sum_i \hat{n}_i(\hat{n}_i-1) - \mu\sum_i \hat{n}_i
\end{equation}
where $\hat{b}^{\dag}_i$ ($\hat{b}_i$) creates (annihilates) an exciton at a site $i$ of the moir\'{e} lattice, while $ \hat{n}_i=\hat{b}^\dag_i \hat{b}_i$ denotes the number operator on the site $i$. Furthermore, $t$  represents the amplitude for tunnelling between two nearest neighbouring sites while $U$ marks the strength of on-site interactions between excitons. Finally $\mu$ provides the chemical potential.  The tunneling $t$ is controlled by the moir\'{e} lattice, i.e. its depth and period, since it is given by the width of the lowest energy band and thereby decreases exponentially with $s$. For a sinusoidal lattice in the limit $V_0\gg E_R$ the tunneling element reads $t=\frac{4}{\sqrt{\pi}}E_R{s}^{3/4}\exp{(-2\sqrt{s})}$. On the other hand, at two dimensions the on-site interaction $U$ is usually expressed as a function of a dimensionless number $\tilde{g}$ and reads $U=\frac{\hbar^2}{4m\pi a_{oh}^2}\tilde{g}$ where $a_{oh}= \sqrt{\frac{\hbar}{m\omega_0}}$, $\omega_0$ being the trapping frequency at the bottom of the moir\'{e} potential within a parabolic approximation \cite{Greiner_PhD}.

For interlayer excitons of TMDs the parameter $\tilde{g}$ quantifying on-site repulsive dipolar interactions has not been measured nor calculated to the best of our knowledge. Nevertheless, we can estimate it relying on experimental and theoretical studies realised in GaAs bilayers. There it was shown that $\tilde{g}\sim5$ \cite{Dang_2019,Dang_2020,Lozovik_QMC} for an exciton dipole moment $d$ equal to 12 nm$\cdot e$ whereas the medium dielectric constant \textcolor{black}{$\epsilon_r$ is around 12.5}. On the other hand, excitons have a dipolar moment of 1 nm$\cdot e$ in TMDs and interact in a medium of dielectric constant of around \textcolor{black}{3}. Accordingly, in TMD bilayers $\tilde{g}$ is approximately 30 times smaller than in GaAs and is then about 0.2. Thus, for a MoSe$_2$/WSe$_2$ bilayer separated by two monolayers of hBN, for $V_0=10$ meV and $a_m=$ 20 nm we find $U\sim$ 0.1 meV whereas $t\sim$ 20 $\mu$eV.

Before proceeding, let us point out that the previous estimation of $\tilde{g}$ is obtained by approximating the dipole-dipole interaction between interlayer excitons by a contact potential. This imposes that the mean spatial separation between excitons is larger than $r_0=m^*d^2/(4\pi\epsilon\hbar^2)$ which characterises the dipolar interaction range (see Supplementals of Ref. \cite{Dang_2019}). For a moir\'{e} lattice with a period of 20 nm and a depth of $5$ meV, using the above values for $d$ and $\epsilon$ we deduce that $r_0\sim$ 7 nm while the full width at half maximum of the lowest energy Wannier function  ($\sim$ 2.35$\cdot a_{oh}$) is around 11 nm in each lattice site. Thus, $\tilde{g}\sim$ 0.2 provides a reasonable estimate for the on-site interaction strength, up to the regime where at most two excitons are confined per site. This corresponds to a maximum density of about  $10^{12}$ cm$^{-2}$. Beyond this value the spatial separation between excitons becomes of the order of $r_0$ and the spatial dependence of the dipolar potential can no longer be neglected. Describing this regime lies beyond the scope of the present work, that is why in the following we restrict our analysis to the situation where at most two excitons occupy individual lattice sites.

\begin{figure}[!ht]
  \includegraphics[width=.9\linewidth]{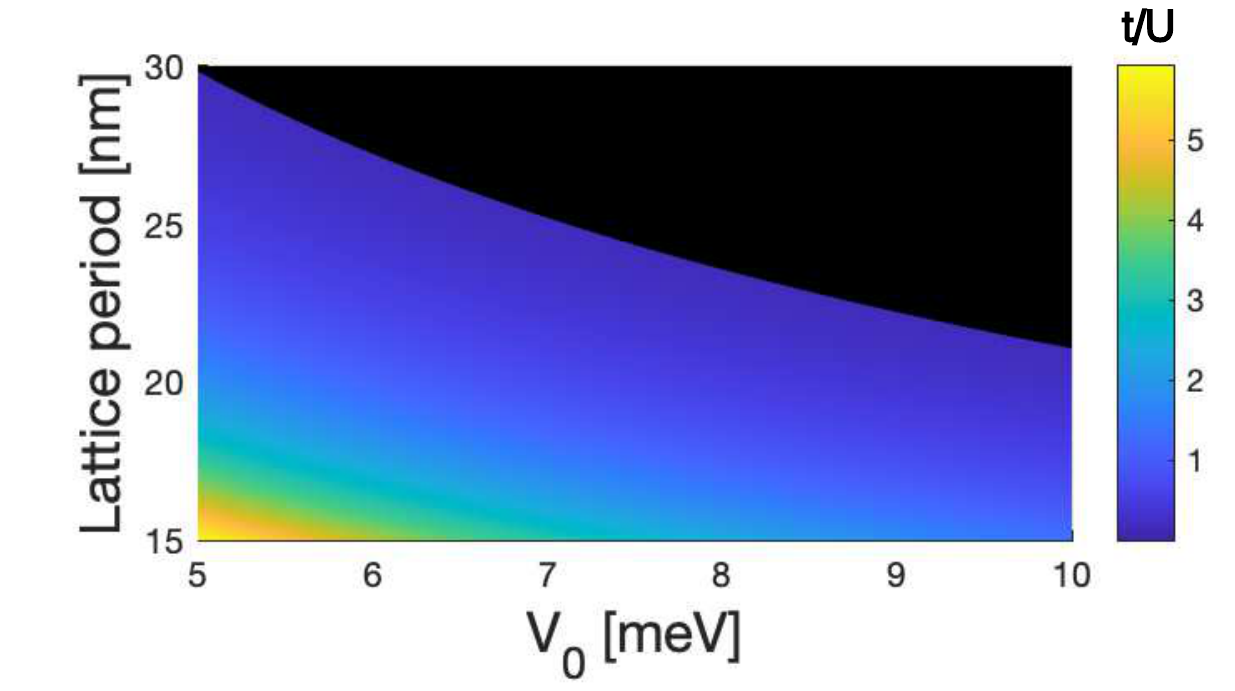}
  \caption{Ratio between tunneling and on-site interaction strengths $t/U$, compared to the critical value  $t/U|_c$ for the buildup of an exciton superfluid phase. The dark region marks the parameter space in which the Mott-insulator phase is energetically more favourable than the superfluid one.}
  \label{fig:fig4}
\end{figure}

The Mott insulator/superfluid transition is possibly characterised simply when one restricts the analysis of the Bose-Hubbard hamiltonian to a single band \cite{Salomon_2010}. First, this approach requires that $V_0\gg E_R$ so that the system evolves in the so-called atomic limit. Then, both the on-site interaction $U$ and the thermal energy need to be small compared to the energy separation between the first two Bloch bands, thus ensuring that only a single band is indeed occupied. \textcolor{black}{For a bilayer device including a hBN spacer we have $s\sim$ $5$. Then only one band is confined (red band diagram in Fig.2.b) and the previous conditions satisfied at low temperatures ($\lesssim$ 30K)}. Accordingly, \textcolor{black}{Eq.(2)} can be treated within a mean-field expansion, i.e. by replacing the original hamiltonian with an effective single-site problem with a self-consistant condition. Thus, one locates the critical coupling for the transition between Mott-insulating and superfluid phases, namely  $\frac{zt}{U}|_c=1/(2N+1+2\sqrt{N(N+1)})$ where $N$ denotes the occupation of lattice sites and $z=3$ is the lattice connectivity \textcolor{black}{(see Ref. \cite{Salomon_2010}, p. 15-20)}. Figure 4 compares $\frac{zt}{U}$ to this critical value for $N=2$. Strikingly, we note that except for the largest periods and lattice depths where the Mott insulator phase is energetically favourable, the superfluid phase is accessible for a broad range of experimental parameters. \textcolor{black}{However, we note that our zero-temperature analysis necessarily overestimates the parameter space where the superfluid phase is favoured, since  at finite temperatures a normal phase separates Mott insulator and superfluid domains \cite{Trivedi_2011}.}

\textit{Discussion} $-$  Two independent experiments have recently reported observations interpreted as manifestations for Bose-Einstein condensation of interlayer excitons. One work was conducted in a MoSe$_2$/WSe$_2$ bilayer without hBN spacer and concluded that quantum degeneracy was reached  at bath temperatures lower than about 8 K for $n\sim 10^{11}$ cm$^{-2}$ \cite{Holleitner_2020}. Importantly, we have shown that in such structure the excitons effective mass is increased between 4$\cdot$10$^3$ and 7$\cdot$10$^5$ for $s$ ranging from 50 to 100. Accordingly, for $n\sim 10^{11}$ cm$^{-2}$ quantum degeneracy is only expected below  900 $\mu$K and 5 $\mu$K respectively. This marks a disagreement as large as 4 to 6 orders of magnitude questioning the role of Bose statistics in the experiments discussed in Ref. \cite{Holleitner_2020}. Another experimental work emphasised a MoSe$_2$/WSe$_2$ bilayer where two monolayers of hBN are intercalated \cite{Shan_2019}. Then, the moir\'{e} lattice is highly screened and the exciton effective mass varied much more weakly. These studies concluded that quantum degeneracy is reached for a density around 3 10$^{11}$ cm$^{-2}$ at 3.5 K. Figure 3 shows that this observation is consistent with our expectations (2 $\lesssim$T$_d\lesssim$ 5K for $n\sim$ 3 10$^{11}$ cm$^{-2}$ and 5 $\lesssim s\lesssim$ 10). Moreover, in Ref. \cite{Shan_2019}  it is argued that quantum degeneracy holds up to 150 K for  $n\sim$ 8 10$^{11}$ cm$^{-2}$. Fig. 3 shows that this conclusion is however out of experimental reach, \textcolor{black}{since $T_d$ is} ranging from 5 to 14 K at this density. Moreover, note that Ref. \cite{Shan_2019} reports that quantum degeneracy is lost when the twist angle between the MoSe$_2$ and WSe$_2$ layer is increased.  Surprisingly this observation goes against our expectations since $s$ decreases for increasing twist angle, so that $T_d$ increases and so does then the degree of quantum degeneracy.

\textcolor{black}{The above discussion questions the role of Bose-Einstein statistics in recent experiments discussing quasi-condensation of interlayer excitons. This may be attributed to the inefficiency of optical probes to signal quantum statistics, since for finite twist angles lowest energy excitons are optically inactive. However, we have emphasised that quasi-condensation occurs in a ground state with finite momentum. This implies that the superfluid phase flows spontaneously, similarly to  the condensation at the roton frequency recently observed with ultra-cold dipolar atoms \cite{Chomaz}. In the superfluid regime we then expect counterflow electron super-currents in the monolayers confining electrons and holes, without any applied in-plane voltage. This behaviour provides an unambiguous signature of exciton superfluidity, in a similar way to quantum Hall bilayers \cite{Nandi_2012} or  twisted bilayer graphene  \cite{PNAS_2011}.}

\textit{Conclusion} $-$  We have highlighted that in TMD bilayers the moir\'{e} potential strongly varies the parameter space where exciton superfluidity is accessible. This is most importantly the case for heterobilayers that are realised without hBN spacer and for which superfluidity is practically out of experimental reach. On the other hand, when hBN is intercalated between the monolayers confining electrons and holes we find that exciton superfluidity is favorable below a critical temperature of slightly less than 10 K for realistic experimental parameters. In fact, the dipolar interaction between excitons limits both the maximum density and the fraction of the superfluid phase. Remarkably, the latter is characterised by a spontaneous flow, which allows one to unambiguously identify the quantum regime. 

\textit{Acknowledgements} $-$ We would like to thank M. Holzmann for stimulating discussions together with M. Polini and B. Urbaszek for a critical reading of the manuscript. Our work has been supported by the Labex Matisse of Sorbonne University and by the ANR contract IXTASE.

\end{document}